\documentstyle[epsfig,12pt,fleqn]{article} 
\begin{document}    
\renewcommand\thepage{\ }
%

%
%
\newcommand\reportnumber{388} 
\newcommand\mydate{October, 1994} 
\newlength{\nulogo} 
\settowidth{\nulogo}{\small\sf{N.U.H.E.P. Report No. \reportnumber}}
\vspace{-1in}
\title{\hfill\fbox{{\parbox{\nulogo}{\small\sf{Northwestern University: \\
N.U.H.E.P. Report No. \reportnumber\\
          \mydate}}}}
          \vspace{.5in} \\
{ A Test of the Standard Model, Using Da$\Phi$ne}
}
\author{Martin~M.~Block
\thanks{
to be published in the Proceedings of the XXIV International Symposium on
Multiparticle
Dynamics, Eds. A.~Giovannini, S.~Lupia and R.~Ugocionni, World Scientific,
Singapore.
Work partially supported by Department of Energy grant
DOE 0680-300-N008 Task B.
} \vspace{-3pt}\\
{\small\em Department of Physics and Astronomy,} \vspace{-5pt} \\ 
{\small\em Northwestern University, Evanston, IL 60208}\vspace{5pt}\\
\vspace{-5pt}\\
%
}
\vfill
\date{} 

\maketitle
\begin{abstract} 
Both the light hypernucleus
${}_\Lambda {\rm H}^4$ and the nucleus ${\rm He}^4$ have spin 0. The ratio $R$
of the electron to muon  rates for the pure Fermi transitions
\begin{equation}
R=\frac{\Gamma\left({}_\Lambda {\rm H}^4\rightarrow {\rm e}^-
+ \bar \nu +{\rm He}^4\right)}
{\Gamma\left({}_\Lambda {\rm H}^4\rightarrow {\mu}^-
+ \bar \nu +{\rm He}^4\right)}
\end{equation}
is a sensitive measure of the presence of a {\em second-class} weak current
(to the extent that SU(3) is valid in strong interactions), and hence, is a
test of the Standard Model. Rates and sensitivities, using Da$\Phi$ne,
the $e^+e^-$ machine  under construction at Frascati, are discussed.
\end{abstract}  
%
%
     \pagenumbering{arabic}
     \renewcommand{\thepage}{-- \arabic{page}\ --}  
%
\section{Introduction}
The ground states of both the light hypernucleus ${}_\Lambda {\rm H}^4$ and
the nucleus ${\rm He}^4$ are spin 0 states\cite{bc}.  Therefore, the weak
decays
\begin{equation}
{}_\Lambda {\rm H}^4\rightarrow {\rm e}^-
+ \bar \nu + {\rm He}^4
\end{equation}
and
\begin{equation}
{}_\Lambda {\rm H}^4\rightarrow {\mu}^-
+ \bar \nu + {\rm He}^4
\end{equation}
are both Fermi transitions, and hence, are both
{\em pure vector transitions}\cite{dick,aps,oakes,betadecay}.
Thus, the hadronic current is
given by the vector current
\begin{equation}
V^\rho =G_\Lambda\gamma^\rho +g_2i\sigma^{\rho\nu}Q_\nu +f_3Q^\rho,
\label{vector}
\end{equation}
where the 4-momentum transfer $Q^\rho$ is given by $\Lambda-p=\ell -\nu$.
The first
term of eq.~(\ref{vector}), $\gamma^\rho$, is the conventional
(and {\em dominant}) term, the second
term, $i\sigma^{\rho\nu}Q_\nu$, is the ``weak magnetism'' term (which
will later neglected as very tiny), both of which are first class
currents, and the third term, $Q^\rho$, is the ``induced scalar'', which
is a second-class current, and hence (to the level that SU(3) is valid
for the strong interactions) is
{\em forbidden} in the Standard Model. Thus, a non-zero $f_3$ would be a
violation of the Standard Model.

This communication will suggest a method of testing for a non-zero $f_3$,
using the Frascati accelerator Da$\Phi$ne.
\section{Experimental Outline}
Da$\Phi$ne, the new $e^+e^-$ machine being constructed at Frascati,
is a copious source of $\Phi$'s, which decay dominantly into
very low energy ${\rm K}^+{\rm K}^-$ pairs.  If one surrounds the
intersection region with a Helium target (either a liquid or a sufficiently
large gaseous target), one can stop the ${\rm K}^-$ and produce the at-rest
reaction
\begin{equation}
{\rm K}^- + {\rm He}^4\rightarrow \pi^0 + {}_\Lambda {\rm H}^4.
\end{equation}
This reaction occurs copiously, being about
1\% of all reactions in which a negative kaon stops in Helium{\cite{bc}.
This hypernucleus production has a unique experimental signature, producing
a monoenergetic and isotropic $\pi^0$ of $\approx 260$ MeV. Subsequently, after
the detection of the $\pi^0$, one can measure the ratio of the electron to
muon rates for
the pure Fermi transition
\begin{equation}
R=\frac{\Gamma({}_\Lambda {\rm H}^4\rightarrow {\rm e}^-
    + \bar \nu + {\rm He}^4)}
	{\Gamma({}_\Lambda {\rm H}^4\rightarrow {\mu}^-
    + \bar \nu + {\rm He}^4)}, \label{R}
\end{equation}
which is a {\em sensitive} measure of $f_3$ in eq.~(\ref{vector}),
{\em i.e.,} the presence of an induced scalar term
(a {\em second-class term})
in the weak hadronic current. It is easy to show that the contraction of
vector term $f_3Q^\rho$ with the leptonic current
${\bar u}_\ell\gamma_\rho(1-i\gamma_5)u_\nu$,
(where $\ell$ stands for lepton, either e or $\mu$), results in an effective
induced scalar term
${\rm m}_\ell f_3\left[{\bar u}_\ell(1-i\gamma_5)u_\nu\right]$, where
the lepton mass  m$_\ell$ is given by either m$_\mu$ or m$_{\rm e}$.
Since the electron mass is so small, the ratio $R$ in eq.~(\ref{R}) is
sensitive to the presence of $f_3Q_\rho$, because the induced scalar
contributes negligibly in the electron channel, whereas it has a strong
influence in the muon channel. The ratio of electrons to muons is thus
very
sensitive to $f_3$, whereas, at the same time, it is very insensitive to
the details of the calculation, since the uncertainties in the nuclear
physics, wave functions, {\em etc.}, is about the same for the muon and the
electron and tends to cancel in the ratio.

Further, in our case,
the only second-class current allowed by
Lorentz invariance is the ``induced scalar'' term $f_3Q^\mu$ , since
{\em only} the vector transition is allowed. The axial
``induced pseudoscalar'' term of the form $Q^\mu\gamma_5$(which
{\em is allowed} in
the free decay of the $\Lambda^0$, and thus `mimics' the
vector term $f_3Q^\mu$) is
{\em not allowed} in the pure vector transition of the hypernucleus. Thus,
we have a unique interpretation of the experimental results.

The remainder of this paper will be devoted
to a quantitative analysis of the ratio $R$, with regard to its sensitivity
to $f_3$.
\section{Decay Rate for ${}_\Lambda {\rm H}^4\rightarrow {\ell}^-
+ \bar \nu + {\rm He}^4$}
The nuclear matrix element ${\cal M}^\rho$ for the $0\rightarrow 0$
transition
${}_\Lambda {\rm H}^4\rightarrow {\ell}^- +\nu + {\rm He}^4$ is
\begin{eqnarray}
\lefteqn{{\cal M}^\rho=\sqrt{2}\int e^{-i(
\vec{\rm p}_\nu \cdot \vec{\rm r}_\nu +\vec{\rm p}_\ell \cdot
\vec{\rm r}_\ell)}
e^{-i\vec{\rm q}\cdot (\vec{\rm r}_p +\vec{\rm r}_2
+\vec{\rm r}_3+\vec{\rm r}_4)]/4}
\Phi_\alpha^*({\rm r}_p,{\rm r}_2;{\rm r}_3,{\rm r}_4)
\chi_f^*({\rm p},2;3,4)\times}
\nonumber \\
&&V^\rho\times\Phi_{\Lambda}({\rm r}_{\Lambda};{\rm r}_2;{\rm r}_3,{\rm r}_4)
\chi_i(\Lambda ;2;,3,4)
\delta({\rm r}_{\rm p}-{\rm r}_{\Lambda})
\delta({\rm r}_\ell-{\rm r}_{\Lambda})
\delta({\rm r}_{\nu} -{\rm r}_{\Lambda})
\delta_{{\rm p}\Lambda}^\sigma\times\nonumber \\
&&\,\,\,\,\,\,\,\,\,\,\,d\tau_{\rm p}\,d\tau_{\Lambda}
\, d\tau_\ell\,d\tau_{\nu}\,
d\tau_2\,d\tau_3\,d\tau_4,\label{nuclearME}
\end{eqnarray}
where $\vec q$, $\vec{\rm p}_{\nu}$ and $\vec{\rm p}_{\ell}$ are the momenta
of the recoil ${\rm He}^4$, the anti-neutrino and the electron (muon), and
$\vec q+ \vec{\rm p}_{\nu} +\vec{\rm p}_{\ell}=0$.  The $\chi$'s are the
spin-wave functions for ${}_\Lambda {\rm H}^4$ and ${\rm He}^4$.  The factor
$\sqrt{2}$ arises from having two identical protons in He$^4$.

For the alpha-particle wave function in eq.~(\ref{nuclearME}), we use
the Gaussian wave function
\begin{equation}
\Phi_\alpha^*({\rm r}_1,{\rm r}_2;{\rm r}_3,{\rm r}_4)=N_4e^{
-\frac{\alpha_4}{2}\sum r_{ij}^2},
\end{equation}
where $\vec r_{ij}=\vec r_j-\vec r_i$, the normalization factor is given by
$N_4=2^{3/4}\left(\frac{2\alpha_4}{\pi}\right)^{9/4}$, and
$\alpha_4=\frac{9}{32R_4^2}$.  The radius $R_4$ is fixed from electron
scattering data, assuming that the matter distribution is the same as
the charge distribution and
allowing for the finite size of the nucleon. We will use $R_4=1.44 f$.
We will assume that the hypernucleus wave function factors into a $\Lambda_0$
moving about a physical triton core, {\em i.e.,}
\begin{eqnarray}
\Phi_{\Lambda}({\rm r}_{\Lambda};{\rm r}_2;{\rm r}_3,{\rm r}_4)&=&
u_\Lambda\left(\left|\vec {\rm r}_{\Lambda}-
\frac{\vec{\rm r_2}+\vec {\rm r}_3+\vec{\rm r}_4}{3}\right|\right)
\Phi_t({\rm r}_2;{\rm r}_3,{\rm r}_4)\nonumber\\
&=&u_\Lambda(\rho)\Phi_t({\rm r}_2;{\rm r}_3,{\rm r}_4)\nonumber\\
&=&u_\Lambda(\rho)N_3e^{-\frac{\alpha_3}{2}\sum r_{ij}^2},\label{hf}
\end{eqnarray}
where in eq.~(\ref{hf}), the triton wave function is assumed to be Gaussian
and $\vec \rho=\vec {\rm r}_{\Lambda}-
\frac{\vec{\rm r}_2+\vec{\rm r}_3+\vec{\rm r}_4}{3}$.
The triton normalization factor is given by
$N_3=\frac{\left(3\alpha_3^2\right)^{3/4}}{\pi^{3/2}}$ and
$\alpha_3=\frac{1}{3R_3^2}$. Again, the radius $R_3$ is fixed from electron
scattering data, assuming that the matter distribution is the same as
the charge distribution and
allowing for the finite size of the nucleon. We use $R_3=1.66 f$.
The wave function $u_\Lambda(\rho)$ has been numerically evaluated by
Dalitz and Downs\cite{dandd}, after solving the Schr\"{o}dinger equation using
a binding energy $\epsilon =1.9$ MeV for the ${}_\Lambda{\rm H}^3$.

Since the calculation of eq.~(\ref{nuclearME}) involves
(non-relativistic) wave functions for helium and the hypernucleus, we
must expand the relativistic hadronic current $V^\rho$ non-relativistically
to get
\begin{eqnarray}
V^\rho&=&\left(G_\Lambda+f_3Q^0,G_\Lambda\left[\vec\sigma
\left(\vec\sigma\cdot\frac{\nabla_{{{\rm r}_{\Lambda}}}}{2m_{\Lambda}}
\right )
-\left(\vec\sigma\cdot\frac{\nabla_{{{\rm r}_{\rm p}}}}
{2m_{\rm p}}\right )
\vec\sigma \right]+f_3\vec Q\right),\label{Vnonrel}
\end{eqnarray}
where the form factors $G^\Lambda$ and $f_3$ are assumed to be functions
of the squared 4-momentum $Q^2$, and are taken to be
\begin{eqnarray}
G_\Lambda(Q^2)&=&G_\Lambda(0)\frac{M^2_{K^*}}{M^2_{K^*}+Q^2},\nonumber\\
f_3(Q^2)&=&f_3(0)\frac{M^2_{K^*}}{M^2_{K^*}+Q^2},
\end{eqnarray}
where $M_{K^*}=894.1$ MeV, the mass of the $K^*$ resonance.
The effect of the form factor
variation will turn out to be very small ($<5$\%). In the non-relativistic
expression for $V^\rho$ in eq.~(\ref{Vnonrel}),
we have neglected completely
the "weak magnetism" term $g_2i\sigma^{\rho\nu}Q_\nu$  of eq.~(\ref{vector}),
completely, which is exceedingly small. As emphasized earlier, the term in
$f_3$ is only important for muon decay, and is negligible for electron decay.
After integration over triton coordinates, using the factorizable wave
functions, we find the nuclear matrix element
\begin{eqnarray}
{\cal M}^0&=&\sqrt{2}{\bar \chi}_f\chi_iF(q)
\left(G_\Lambda + f_3Q^0\right),\nonumber\\
\vec{\cal M}&=&\sqrt{2}{\bar \chi}_f\chi_iF(q)
\,\left\{G_\Lambda (\vec q\cdot{\vec \sigma}_1)
{\vec \sigma}_1\frac{1+1.366}{2m_4}
+f_3\vec Q\right\},
\end{eqnarray}
where $F(q)$, the nuclear overlap integral is
\begin{equation}
F(q)=\left[ \frac{48\alpha_3 \alpha_4}{(3\alpha_3+\alpha_4)^2}\right]^{3/2}
G(q),
\end{equation}
with
\begin{eqnarray}
\lefteqn{G(q)=(3\alpha_4/\pi)^{3/4} \int \exp (-\frac{3}{2}\alpha_4
\rho^2)}\nonumber\\
&&\times \exp \left [ i\frac{3}{4}\vec{q}\cdot\vec{\rho}\right ]
u_{\Lambda}(\rho)\,d\rho,
\end{eqnarray}
and $m_4\equiv 4m$.  Dalitz and
Downs have evaluated G(q) numerically, as a function of q, using the
numerical wave function $u_{\Lambda}(\rho)$, for the
binding energy
$\epsilon ({}_{\Lambda}{\rm H}^4)=1.9$ MeV.  In the correction term
proportional to
$\frac{1}{2m_4}$, the term $\frac{1.366}{2m_4}$ is due to the nuclear
corrections arising from the nuclear wave functions employed. We have also
substituted $m_4=4m_p$ in the above correction terms. The quantity $F^2(q)$
is recognized as the ``sticking probability'' that the decay proton from the
$\Lambda_0$ inside the hypernucleus and the triton core of the hypernucleus
overlap to form ${\rm He}^4$.

The total matrix element $M$ is given by
\begin{equation}
M=\frac{1}{\sqrt{2}}\left({\cal M}^0j_0-\vec{\cal M}\cdot\vec j\right),
\end{equation}
where the lepton current is given by
\begin{equation}
j_\rho=(j_0,\vec j)={\bar u}_\ell\,\gamma_\rho(1-i\gamma_5)\,u_\nu.
\end{equation}

The squared matrix element, summed and averaged over spins, for the
reaction\linebreak%
${}_\Lambda {\rm H}^4\rightarrow {\rm e}^- + \bar \nu + {\rm He}^4$, is given
by
\begin{eqnarray}
\lefteqn{\frac{1}{2J+1}\left(m_\ell m_\nu\sum_{\scriptscriptstyle
{\rm spins}} \,|M|^2\right)}
&&\nonumber\\
   &=&2F^2(q^2)\times
\left\{
	{|G_{\Lambda}}|^2 \left[
	(E_{\nu}(E_\ell+p_\ell x)\left(1+2.37\frac{\Delta m}{m_4}\right)
	-2.37\frac{{m_\ell}^2E_\nu}{m_4}\right]\right.\nonumber\\
&&+\left.2\Re (G_\Lambda^*f_3)\left[m_\ell^2E_\nu
			\left\{
			 1+2.37\frac{E_\nu+p_\ell x}{2m_4}
				\right\}\right]
+|f_3|^2\left[ m_\ell^2E_\nu(E_\ell-p_\ell x)\right]\right\}
\end{eqnarray}
where $x={\hat{p}}_\ell\cdot {\hat{p}}_{\nu}$,
${\vec{p}}_\ell+{\vec{p}}_{\nu}+\vec{q}=0$,
$\Delta m$ is the mass difference $m_{{\rm He}^4}-m_{{}_\Lambda {\rm H}^4}$,
and
$m_4=4m_p$, where $m_p$ is the nucleon (proton) mass.

The correction term $[1+2.37(\Delta m/m_4)]$ has been evaluated using the
approximate energy relation $\Delta m\approx E_\ell+E_{\nu}$, and the form
factor can be adequately numerically approximated by
\begin{equation}
F^2(q^2)=a-bq^2+cq^4-dq^6, \label{formfactor}
\end{equation}

with $q$ in units of 100 MeV/c, and
$a=0.723,b=0.170,c=0.0166,d=0.0025$ (taken
from Dalitz and Downs\cite{dandd}.

We consider the motion of the ${\rm He}^4$ to be
non-relativistic.
Thus, the energy and momentum conservation conditions are
\begin{eqnarray}
\Delta m&=&E_\ell+E_{\nu}+\frac{q^2}{2m_{{\rm He}^4}}, \label{energy}\\
0&=&{\vec{p}}_\ell+{\vec{p}}_{\nu}+\vec{q}, \label{momentum}
\end{eqnarray}
where $m_{{\rm He}^4}$, the mass of the ${\rm He}^4$, will be approximated
by $m_4$.
We obtain
$q^2=p_\ell^2+E_{\nu}^2+2p_\ell E_{\nu}x.$
In terms of the lepton energy $E_\ell$, momentum $p_\ell$ and the
mass difference ${\Delta m}$,
we get
\begin{equation}
E_{\nu}=\Delta m - E_\ell-
 \frac{p_\ell^2+(\Delta m-E_\ell)^2 +2p_\ell
  (\Delta m -E_\ell )x}{2m_4},
\end{equation}
using the {\em approximate} energy conservation relation
$E_\ell=\Delta m -E_{\nu}$ in the
small correction term proportional to $1/m_4$. Energy
conservation limits the electron (muon) energy to
\begin{equation}
m_\ell\le E_\ell \le E_{\rm max},
\end{equation}
where
\begin{equation}
E_{\rm max}=\Delta m( 1 - \frac{\Delta m}{2m_4})+\frac{m_\ell^2}{2m_4}
(1-\frac{\Delta m}{m_4}).
\end{equation}
To obtain the phase space needed for the decay process, it useful to find
the quantity $\frac{\partial E_{\nu}}{\partial \Delta m}$, which is given by
\begin{equation}
\frac{\partial E_{\nu}}{\partial \Delta m}=1-
  \frac{\Delta m - E_\ell+p_\ell x}{m_4}.
\end{equation}
In order to find the lepton energy spectrum, $d\Gamma /dE_{\ell}$, we must
integrate over phase space the quantity
\begin{eqnarray}
{d^9}\Gamma& =& (2\pi)^4\,
\frac{1}{2J+1}\left(m_\ell m_\nu{\displaystyle\sum_{\scriptscriptstyle
{\rm spins}}}|M|^2\right)\,
\frac{d^3 {\vec p}_\ell}{(2\pi)^3E_\ell}
\,\frac{d^3 {\vec p}_\nu}{(2\pi)^3E_{\nu}}
\,\frac{d^3 {\vec q}}{(2\pi)^3}\times   \nonumber \\
&&\delta ^3( {\vec{p}}_\ell+{\vec{p}}_{\nu}+\vec{q})
\delta \! \left( \Delta m-( E_\ell+E_{\nu}+\frac{q^2}{2m_4})
    \right),
\end{eqnarray}
where  non-relativistic kinematics for the recoiling ${\rm He}^4$ nucleus,
including the substitution of $m_4$ for $m_{{\rm He}^4}$, have been used.
Using $J=0$, after
integrating over $p_{\nu}$,
as well as over ${\vec q}$,  using
energy and momentum conservation,
we obtain
\begin{eqnarray}
{d^5}\Gamma& =&{\displaystyle\frac{1}{(2\pi)^2}}\,\left(m_\ell m_\nu
{\displaystyle\sum_{\scriptscriptstyle {\rm spins}}}|M|^2\right)
\,\frac{d^3 {\vec p}_e}{(2\pi)^3E_e}
\, p_e E_{\nu}\,\frac{\partial E_{\nu}}{\partial \Delta m}
\,dx.
\end{eqnarray}
We next substitute
for $\partial E_{\nu}/\partial \Delta m$  and integrate over the azimuthal
angle $\phi_{\ell\nu}$
($\int d\phi_{\ell\nu}=2\pi$) to get,
using the relation $p_\ell\,dp_\ell=E_\ell\,dE_\ell$,
\begin{eqnarray}
{d^4}\Gamma& =&{\displaystyle\frac{1}{(2\pi)^4}}\,\left(m_\ell m_\nu
{\displaystyle\sum_{\scriptscriptstyle {\rm spins}}}|M|^2\right)
\, (1-\frac{\Delta m - E_\ell+p_\ell x}{m_4})\, p_\ell E_{\nu}
 \,dx\,dE_\ell \,d\Omega_\ell.\label{lepspectrum}
\end{eqnarray}
Using $q^2={E_\ell}^2+{E_{\nu}}^2 +2p_\ell E_{\nu}x$, and substituting for
$E_{\nu}$ the relation (in terms of $E_\ell , p_\ell$ and $x$), and using
$\int \int d\Omega_\ell=4\pi$, we can
rewrite the lepton (electron or muon) spectrum $\frac{d\Gamma}{dE_\ell}$ as
\begin{eqnarray}
\frac{d\Gamma}{dE_\ell}& =&\frac{1}{4\pi^3} \,
\int^{+1}_{-1}\left(m_\ell m_\nu {\displaystyle\sum_{\scriptscriptstyle
{\rm spins}}}|M|^2
\right)\,
\left((1-\frac{\Delta m - E_\ell+p_\ell x}{m_4}\right)p_\ell E_\nu\,dx.
\label{rate1}
\end{eqnarray}
The integration over $x$ was performed analytically, using {\em Mathematica}.
The total rate for both muons and electrons was found by integrating
Eq.~(\ref{lepspectrum})
numerically, from $m_\ell\le E_\ell \le E_{max}$. The energy spectra were then
converted into momentum spectra, using the relation
$\frac{d\Gamma}{dp_\ell}=\frac{d\Gamma}{dE_\ell}\frac{p_\ell}{E_\ell}$.
The momentum spectra, for the case of $f_3=0$, {\em i.e.,} for the case of
{\em no} second-class currents, are plotted in Fig.~1, with the electron
spectrum being the full line and the muon spectrum being the dashed line.
Both of these spectra have been normalized to unit
area.
\begin{figure}[htb]
 \centerline{\psfig{figure=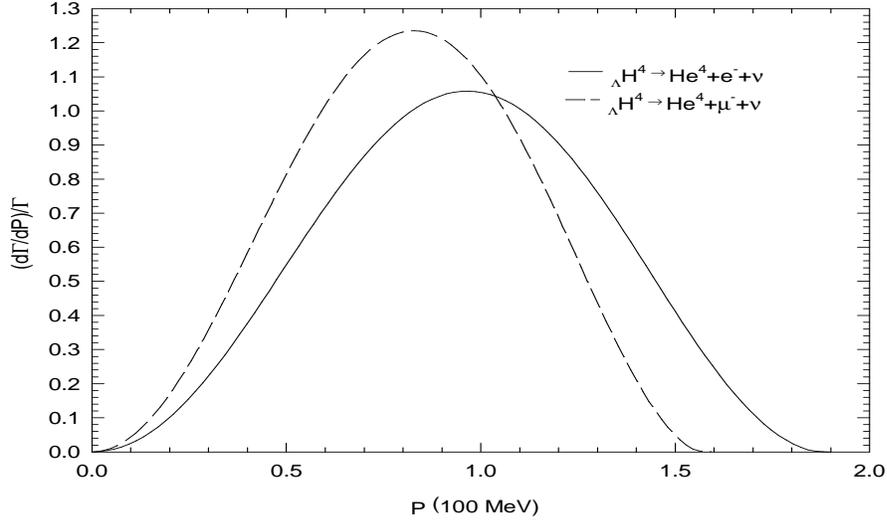,height=5in,width=5in}}
\vspace{-1.5in}
\caption{\protect\small The normalized lepton momentum spectrum
$\frac{1}{\Gamma}\,\frac{d\Gamma}{dp}$ {\em vs.} the lepton momentum
$p$, in units of 100 MeV.
The solid curve is for the electron and the
 dashed curve is for the muon.}
\end{figure}

In absolute units, the spectra have been integrated, using
$G_\Lambda(0)=\sqrt{\frac{3}{2}}G_F\sin \theta_C$, where the Fermi coupling
constant $G_F$ is given by $\frac{1.01\times 10^{-5}}{m_p^2}$ and $\theta_C$,
the Cabbibo angle, was taken as 0.26. The results for $f_3=0$
(no second-class current) are
\begin{eqnarray}
\Gamma_e&=&2.18\times 10^6 {\rm sec}^{-1},\nonumber \\
\Gamma_\mu&=&0.742\times 10^6 {\rm sec}^{-1}.\label{rates}
\end{eqnarray}
\section{Experimental Predictions}
Using a lifetime\cite{bc,piplus2} for the ${}_\Lambda$H$^4$ of
$2\times 10^{-10}$ sec., and with the rates of eq.~(\ref{rates}), we find
{\em branching ratios} for the electron and muon decays of
$R_e=\frac{\Gamma_e}{\Gamma_{\rm all}}=4.36\times 10^{-4}$ and
$R_\mu=\frac{\Gamma_\mu}{\Gamma_{\rm all}}=1.48\times 10^{-4},$
respectively. Using a production rate\cite{bc} of $10^{-2}$
${}_\Lambda$H$^4$
per stopped K$^-$ in He$^4$, we find, for $!0^9$ stopped K$^-$ in a
helium target, that 4600 decays of ${}_\Lambda {\rm H}^4\rightarrow
{\rm e}^-
    + \bar \nu + {\rm He}^4$ and
1570  decays of ${}_\Lambda {\rm H}^4\rightarrow {\mu}^-
    + \bar \nu + {\rm He}^4$ are expected.  These results are summarized in
Table I, where
we have used the production rates and branching ratios of
${}_\Lambda {\rm H}^4\rightarrow
\mbox{{\rm all e}}^-$,
${}_\Lambda {\rm He}^4\rightarrow {\rm \mbox{all e}}^-$ and
${}_\Lambda {\rm H}^3\rightarrow \mbox{{\rm all e}}^-$
taken from ref. \cite{betadecay} to calculate these event rates.

In Table II, we summarize the sensitivity of the experiment to assumed values
of $f_3$, the amplitude that {\em violates} the Standard Model. There,
the amplitude $f_3$ has units of inverse mass $m_\mu^{-1}$. If instead,
we reexpress the amplitude as $f_3=\frac{1}{\cal M}$, we see from Table II
that we can reach a limit of
${\cal M} \approx 10$ GeV for an experimental sensitivity of
2\%. This is also the level where we might expect SU(3) violations to play
a role.

\newpage
\begin{center}
{\Large Table I}
\vspace{.3in}
\end{center}
\begin{center}
\large{Summary---Experimental Rates for Hyperfragment $\beta$-Decay}
\end{center}
\begin{tabbing}
For \=$10^9$ stopped ${\rm K}^-$ in a ${\rm He}^4$ target, we expect:\\
\\
\> 7600 decays of ${}_\Lambda {\rm H}^4\rightarrow \mbox{{\rm all e}}^-$,\\
\> 10,500 decays of ${}_\Lambda {\rm He}^4\rightarrow {\rm
\mbox{all e}}^-$,\\
\\
\> 1200 cases of ${}_\Lambda {\rm H}^3\rightarrow \mbox{{\rm all e}}^-$,\\
\> of which 400 are
${}_\Lambda {\rm H}^3\rightarrow {\rm e}^-
    + \bar \nu + {\rm He}^3$.
\\
\\
\\
\hfill\hspace{2.3in}{\large Pure Fermi Transitions}\hfill
\\
\\
\>4600 ${}_\Lambda {\rm H}^4\rightarrow {\rm e}^-
    + \bar \nu + {\rm He}^4$, \\
\>1570  ${}_\Lambda {\rm H}^4\rightarrow {\mu}^-
    + \bar \nu + {\rm He}^4$, \\
\> \\
\end{tabbing}
\vspace{.1in}
\begin{center}--------------------------------------------%
-------------------------------------------------------------------------
\end{center}
\vspace{.3in}
\begin{center}
{\Large Table II}
\vspace{.15in}
\end{center}
\begin{center}
\large{---The Standard Model Test---\\
$R$, the Electron to Muon Ratio as a Function of the Form Factor $f_3$}
\end{center}
\vspace{.3in}
We express the form factor $f_3$ in units of the inverse muon mass
$m_\mu^{-1}$.\\
$f_3$ ($m_\mu$) \hfill $R=\frac{\Gamma\left({}_\Lambda {\rm H}^4\rightarrow
{\rm e}^-
+ \bar \nu +{\rm He}^4\right)}
{\Gamma\left({}_\Lambda {\rm H}^4\rightarrow {\mu}^-
+ \bar \nu +{\rm He}^4\right)}
$
\\
1.0 \hfill 0.72\\
0.1 \hfill 2.46\\
0.01 \hfill 2.88\\
0 \hfill 2.94\\
-.01 \hfill 2.99\\
-.1\hfill 3.52\\
-1.0\hfill 5.50
\\Summary: At the level of 2\%, we are sensitive to a mass scale of
$\approx$ 10 GeV.

\begin{thebibliography}{99}
\bibitem{bc} M.~M.~Block {\em et al.}, {\em Proceedings of the Tenth
Annual
International Conference on High Energy Physics}, edited by
E.~C.~G.~Sudarshan, J.~H.~Tinlot, and A.~C.~C.~Melissinos (Interscience
Publishers, Inc. New York, 1961).
\bibitem{dick}The significance of this $0\rightarrow 0$ transition
was originally
pointed out by R.~H.~Dalitz, Proceedings of the International
Conference on
Hyperfragments, St. Cergue, 1963, CERN Report No. 64-1, 1964, p. 206
\bibitem{aps}M.~M.~Block, Bull. Am. Phys. Soc. {\bf 10}, 1105 (1965).
\bibitem{oakes}P~McNamee and R.~J.~Oakes, Phys. Rev. {\bf 149}, 1157
(1966).
\bibitem{betadecay}M.~M.~Block, Phys. Rev. {\bf 168}, 1795 (1968).
\bibitem{dandd}R.~H.~Dalitz and B.~W.~Downs, Phys Rev. {\bf 111}, 967
(1958).
\bibitem{cabbibo}N.~Cabbibo, Phys. Rev. Letters {\bf 10}, 531 (1963).
\bibitem{piplus2}G.~Keyes {\em et al.},
Nuovo Cimento {\bf 31 A, N. 3},
491 (1976).
\end{thebibliography}
\end{document}